\documentstyle[12pt,a4,amssymb]{article}
\newcounter{foot-ctr}

\newmathalphabet*{\mbf}{cmr}{b}{n}
\newmathalphabet*{\mii}{cmr}{m}{it}
\newmathalphabet*{\mtt}{cmtt}{m}{n}

\renewcommand{\b}{{\mathrm b}}
\renewcommand{\c}{{\mathrm c}}

\newcommand{\p}{{\mathrm p}}

\newcommand{\J}{{\mathrm J}}

\newcommand{\pbar}{\overline{\mathrm p}}

\newcommand{\ppbar}{\p\pbar}

\newcommand{\JP}{\J/\psi}

\begin{document}
\sloppy

\pagestyle{empty}

\begin{flushright}
CERN-TH.7170/94\\
\end{flushright}
\
\vskip1.5cm
\begin{center}
  {\LARGE\bf Quarkonium production and decays$^a$}
\end{center}
\bigskip
\begin{center}
{\Large Gerhard A.\ Schuler} \\[3mm]
{Theory Division, CERN, CH-1211 Geneva 23, Switzerland}\\
\end{center}
\vskip1.5cm
\noindent
\baselineskip=14pt \noindent
{\bf Abstract}

\noindent
Quarkonium decays are studied in the charmonium model.
Relativistic corrections, higher-order perturbative QCD corrections
and non-perturbative contributions are discussed.
Recent measurements of charmonium annihilation rates are used
to evaluate the strong coupling constant $\alpha_s$ simultaneously
with the wave functions (and their derivatives) at the origin.
Further predictions are made for yet unobserved decay rates.
The various models for quarkonium production in hadronic
collisions are critically reviewed. Based on the charmonium
model, the cross sections of different quarkonium states
are given in a well-defined QCD perturbation series,
including quark--antiquark, quark--gluon, and gluon--gluon scatterings.
Numerical estimates are given for charmonium production in
$\p\p$, $\ppbar$, and $\pi\p$ collisions. The r{\^o}le of
indirect $\JP$ production via $\chi_{\c J}(1P)$, $\eta_{\c}(2S)$,
$\psi(2S)$ and $\b$-decays is pointed out. Relativistic effects and
non-perturbative contributions are found to be important.
Existing measurements are compiled and shown to be well
explained if all contributions are included. The ${}^1S_0$ cross section
is calculated in complete next-to-leading order.
Finally, a study of the high-energy behaviour of quarkonium cross
sections is made, based on the asymptotical behaviour of higher-order
QCD corrections.

\vskip1.0cm
\begin{center}
 {\it submitted to Physics Reports}
\end{center}

\vspace{\fill}
\noindent
\rule{6cm}{0.4mm}

\vspace{3mm} \noindent
{\Large {\bf $^a$}} Habilitationsschrift, Univ.\ of Hamburg, Germany, 1993

\vspace{\fill}
 \noindent
CERN-TH.7170/94 \\
February 1994

\clearpage
\pagestyle{plain}
\pagenumbering{roman}
\setcounter{page}{1}
%\addtolength{\textheight}{+40mm}
\section*{Contents}
\contentsline {section}{\numberline {1}Introduction}{1}
\contentsline {section}{\numberline {2}
Models of quarkonium hadroproduction}{6}
\contentsline {subsection}{\numberline {2.1}Drell--Yan process}{6}
\contentsline {subsection}{\numberline {2.2}Direct
${\mathrm J}/\psi $ production via ${\mathrm q}\overline
{\mathrm q}$ and ${\mathrm c}\overline {\mathrm c}$ annihilation}{8}
\contentsline {subsection}{\numberline {2.3}The r{{\accent 94 o}}le
of gluons: $\eta _{{\mathrm c}}$ production via gluon fusion}{10}
\contentsline {subsection}{\numberline {2.4}Duality approach}{12}
\contentsline {subsection}{\numberline {2.5}Indirect ${\mathrm J}/\psi $
production}{12}
\contentsline {subsection}{\numberline {2.6}Quarkonia production in
perturbative QCD}{14}
\contentsline {subsection}{\numberline {2.7}A first comparison with
data}{17}
\contentsline {section}{\numberline {3}Duality approach}{19}
\contentsline {subsection}{\numberline {3.1}The model}{19}
\contentsline {subsection}{\numberline {3.2}Assumptions and limits of
validity}{20}
\contentsline {subsection}{\numberline {3.3}An improved duality
approach}{22}
\contentsline {section}{\numberline {4}Charmonium model}{24}
\contentsline {subsection}{\numberline {4.1}The idea}{24}
\contentsline {subsection}{\numberline {4.2}The potential}{25}
\contentsline {subsection}{\numberline {4.3}Corrections and
limitations}{26}
\contentsline {section}{\numberline {5}Quarkonium decays}{28}
\contentsline {subsection}{\numberline {5.1}${}^1S_0$ decays}{29}
\contentsline {subsection}{\numberline {5.2}$n{}^3S_1$ decays}{34}
\contentsline {subsection}{\numberline {5.3}Relativistic corrections}{40}
\contentsline {subsection}{\numberline {5.4}$P$-decays}{43}
\contentsline {section}{\numberline {6}Experimental results on
${\mathrm J}/\psi $, $\psi (2S)$, and $\chi _{{\mathrm c}J}$
hadro\discretionary {-}{}{}pro\discretionary {-}{}{}duction}{48}
\contentsline {subsection}{\numberline {6.1}Kinematics}{48}
\contentsline {subsection}{\numberline {6.2}${\mathrm J}/\psi $ cross
sections}{49}
\contentsline {subsection}{\numberline {6.3}$\chi _{{\mathrm c}J}$ and
$\psi (2S)$ cross sections and the components of ${\mathrm J}/\psi $
production}{55}
\contentsline {section}{\numberline {7}Quarkonium production in the
charmonium model}{58}
\contentsline {subsection}{\numberline {7.1}Assumptions}{58}
\contentsline {subsection}{\numberline {7.2}${}^1S_0$ production}{60}
\contentsline {subsection}{\numberline {7.3}${}^3S_1$ cross sections}{64}
\contentsline {subsection}{\numberline {7.4}$\chi _{J}$ production}{66}
\contentsline {subsection}{\numberline {7.5}Relativistic corrections
to ${}^3S_1$ and ${}^3P_1$ production}{69}
\contentsline {section}{\numberline {8}Cross sections of quarkonium
hadroproduction}{73}
\contentsline {subsection}{\numberline {8.1}${}^1S_0$ cross sections}{74}
\contentsline {subsection}{\numberline {8.2}${}^3P_{0,2}$ cross
sections}{75}
\contentsline {subsection}{\numberline {8.3}${}^3L_{1}$ cross
sections}{76}
\contentsline {subsection}{\numberline {8.4}Inclusive ${\mathrm J}/
\psi $ cross section and the contributions from higher states}{77}
\contentsline {subsection}{\numberline {8.5}Pion-induced cross
sections}{79}
\contentsline {subsection}{\numberline {8.6}Toponium production at LHC
energies}{82}
\contentsline {subsection}{\numberline {8.7}High-energy extrapolations}
{83}
\contentsline {section}{\numberline {9}Conclusions and outlook}{109}
\end{document}